\documentstyle[aps,psfig,amsmath]{revtex}
\newcommand{\be}{\begin{equation}}
\newcommand{\ee}{\end{equation}}
\newcommand{\bea}{\begin{eqnarray}}
\newcommand{\eea}{\end{eqnarray}}
\newcommand{\bef}{\begin{figure}}
\newcommand{\eef}{\end{figure}}
\newcommand{\bm}{\bibitem}
\newcommand{\al}{\alpha}
\newcommand{\bet}{\beta}
\newcommand{\lm}{\lambda}
\newcommand{\sg}{\sigma}
\newcommand{\de}{\delta}

\newcommand{\ep}{\epsilon}
\newcommand{\gamu}{\gamma_{\mu}}
\newcommand{\gamU}{\gamma^{\mu}}
\newcommand{\om}{\omega}
\newcommand{\rw}{\rightarrow}
\newcommand{\mn}{\mu\nu}
\newcommand{\cl}{{\cal{L}}}
\newcommand{\br}{{\boldsymbol{\rho}}}
\newcommand{\bp}{{\boldsymbol{\pi}}}

\newcommand{\bP}{{\boldsymbol{\Pi}}}
\newcommand{\ba}{{\boldsymbol{a}}}
\newcommand{\bv}{{\boldsymbol{v}}}
\newcommand{\D}{{\boldsymbol{D}}}
\newcommand{\U}{{\boldsymbol{U}}}
\newcommand{\R}{{\boldsymbol{R}}}
\newcommand{\F}{F_\pi}
\newcommand{\Fr}{F_\rho}
\newcommand{\dmd}{\partial_\mu}
\newcommand{\dmu}{\partial^\mu}
\newcommand{\dnd}{\partial_\nu}
\newcommand{\dnu}{\partial^\nu}
\newcommand{\emn}{\epsilon_{\mu\nu\lambda\sigma}}
\newcommand{\rt}{\sqrt 2}

\newcommand{\fmn}{f^{\mn}}
\newcommand{\vpamu}{v_\mu+a_\mu}
\newcommand{\vmamu}{v_\mu-a_\mu}

\newcommand{\omk}{\omega_k}

\newcommand{\bq}{\bar{q}}
\newcommand{\vk}{\vec{k}}
\newcommand{\vq}{\vec{q}}
\newcommand{\la}{\langle}
\newcommand{\ra}{\rangle}

\newcommand{\ta}{\tau^a}
\begin{document}

\setcounter{page}{1}

\title{Vector and axial-vector mesons at finite temperature}

\author{S. Mallik} 
\address{Saha Institute of Nuclear Physics,
1/AF, Bidhannagar, Kolkata-700064, India} 

\author{Sourav Sarkar} 
\address{Variable Energy Cyclotron Centre, 1/AF, Bidhannagar, Kolkata-700064,
 India}


\maketitle

\begin{abstract} 

We consider the  thermal correlation functions of vector and 
axial-vector currents and evaluate corrections to the vector and
axial-vector meson pole terms to one loop in chiral perturbation
theory. As expected, the pole positions do not shift to leading order 
in temperature. But the residues decrease with temperature.
 
\end{abstract}


\section{Introduction}

An important topic in strong interaction at non-zero temperature is the
calculation of shifts in the masses and the couplings which hadrons suffer 
while moving through
a heat bath. Apart from their immediate relevance in understanding the
experimental data on heavy ion collisions, these shifts as functions of
temperature provide an effective description of the thermal properties of
strong interaction. Mention must be made here also of the condensates, which
are related more directly to the condensed matter aspects of strong
interaction, particularly to the possibility of phase transition with the
rise of temperature. 

At low energies processes involving strong interaction as described by QCD
are best analysed in chiral perturbation theory \cite{Gasser84}. 
It has long been established phenomenologically that the symmetry group 
$SU(3)_R\times SU(3)_L$ of the QCD Lagrangian of three flavours of massless 
quarks is broken spontaneously to its diagonal subgroup $SU(3)_V$, giving
rise to the octet of Goldstone bosons. Chiral perturbation theory 
realises this symmetry nonlinearly on the physical fields. This framework 
restricts considerably the interactions among hadrons involving the Goldstone
bosons. It has led to a whole set of accurate analysis of low
energy hadronic processes \cite{Gasser85,Ecker95}.

Chiral perturbation theory finds a natural application to problems of strong
interaction at non-zero temperature $T$. At low temperature pions dominate the
heat bath, the other hadrons being exponentially suppressed. 
One is thus led to consider the reduced chiral symmetry group  
$G=SU(2)_R\times SU(2)_L$, which breaks spontaneously to the isospin
subgroup $H=SU(2)_V$. This 
symmetry implies that interactions involving low energy pions are weak. In
calculating quantities at low temperature these interactions can
therefore  be treated as perturbation. A wide variety of problems has been
treated in this way \cite{Gasser87,Gasser87a}. In particular, it has been 
applied to 
the correlation function of nucleon currents to find the position of 
the nucleon pole and its residue at finite temperature \cite{Leutwyler90}. 
In this work we apply a similar procedure
to the isotriplets of vector and axial-vector mesons. 

Here we consider the correlation functions of vector and axial-vector
currents defined in QCD. To calculate the pole parameters to $O(T^2)$, 
we draw all the one loop Feynman diagrams with vector and axial-vector
mesons and pions, which contribute to these correlation functions. 
With the form of
interaction vertices prescribed by chiral perturbation theory, it is simple 
to evaluate the finite temperature parts of these diagrams near the poles in
the chiral limit. Although we begin with a large number of diagrams, only a
few actually contribute to the meson poles to order $T^2$. We also argue
that other particles not included in our scheme cannot change the results to
this order. 

In sec. II we summarise the chiral transformation rules of the relevant
field variables to arrive at the chirally invariant 
Lagrangian of the spin one mesons and pions. In sec. III we obtain explicitly
the interaction vertices needed to evaluate the Feynman diagrams. In sec. IV 
we find the shift in the masses and couplings of these mesons to $O(T^2)$. 
Finally in sec. V we summarise our results and comment on other attempts to 
determine these shifts. The Appendix discusses the $2\times 2$ matrix 
and the Lorentz tensor structures of the correlation
functions in the real time thermal field theory.

\section{Chiral Perturbation Theory} 
\setcounter{equation}{0}
\renewcommand{\theequation}{2.\arabic{equation}}

It is convenient to review in this section the 
definition of the appropriate field
variables and their transformation properties, leading to the effective
Lagrangian of chiral perturbation theory \cite{Ecker89,Ecker89a}. 
The Goldstone fields $\pi^a(x)$ reside in the coset space of $G$ with 
respect to $H$. A standard parametrization of this space allows one to 
collect these  fields in the form of an unitary matrix,
\be
u(x) = e^{i\pi (x)/{2\F}}~,~~~~~~~~\pi (x)=\sum_{a=1}^3\pi^a(x)\ta~~,
\ee
where the constant $F_\pi$ can be identified with the pion decay constant, 
$\F =93$ MeV and $\ta$ are the Pauli matrices. 
The matrix  $u$ transforms under $G$ according to the rule,
\be
u\rw g_R u h^\dag=hug_L^\dag~~,
\ee
where $g_{R,L}\in SU(2)_{R,L}$ and $h(\pi)\in SU(2)_{V}$.
The matrix $U=u^2$ then transforms as 
\be
U\rw g_R\,U\,g_L^\dag.
\ee
Any multiplet $\psi(x)$ of non-Goldstone fields transforms as
\be
\psi\rw h\psi,
\ee
with $h$ in the appropriate representation. In particular, if they also
belong to the adjoint (triplet) representation, such as the vector and the 
axial-vector mesons we are considering here, we may use the same $h$ as 
in (2.2): Denoting the triplet fields by $R_\mu(x)$, 
\be
R_\mu (x)=\frac{1}{\sqrt 2}\sum_{a=1}^3\,R_\mu^a (x)\ta, 
\ee
it transforms as 
\be
R_\mu\rw h\,R_\mu\,h^\dag.
\ee
The singlet field $S_{\mu}(x)$, of course, remains unchanged,
\[ S_{\mu} \rw S_{\mu} .\]

As already stated we are concerned here with the evaluation of the
two-point functions of vector and axial-vector currents of QCD,
\be
V_\mu^a(x)=\bq(x)\gamu\frac{\ta}{2}q(x),~~~~~~~~~
A_\mu^a(x)=\bq(x)\gamu\gamma_5\frac{\ta}{2}q(x),
\ee
in the effective field theory. This is most conveniently carried out in the 
external field method, where one introduces external fields $v_\mu^a(x)$ and
$a_\mu^a(x)$ coupled to the currents $V_\mu^a(x)$ and 
$A_\mu^a(x)$ \cite{Gasser84}. The original QCD Lagrangian 
${\cal L}^{(0)}_{QCD}$ of massless quarks is then augmented to 
\bea
&&{\cal L}^{(0)}_{QCD}+v_\mu^a(x)V^\mu_a(x)+a_\mu^a(x)A^\mu_a(x)\nonumber\\
&=&i\bq_R\gamU\{\dmd-i(\vpamu)\}q_R+
i\bq_L\gamU\{\dmd-i(\vmamu)\}q_L+\cdot\cdot\cdot
\eea
where $q_R (q_L)$ is the right (left) handed component of $q$, 
$q_{R,L}= \frac{1}{2} (1\pm\gamma_5)q$.  
The ellipsis denote the (flavour neutral) gauge field term,
and $v_\mu(x)$ and $a_\mu(x)$ are matrices in flavour space,
\be
v_\mu(x)=v_\mu^a(x)\frac{\ta}{2}~~~,~~~~~~
a_\mu(x)=a_\mu^a(x)\frac{\ta}{2}.
\ee
The global invariance of ${\cal L}^{(0)}_{QCD}$ is now raised to local
invariance under the transformations
\bea
q_R^\prime&=&g_R q_R~~~, ~~~~~
q_L^\prime=g_L q_L,\nonumber\\
v_\mu^\prime\pm a_\mu^\prime &=&g_{R,L}(v_\mu\pm a_\mu)g_{R,L}^\dag
+ig_{R,L}\dmd g_{R,L}^\dag,
\eea
where the group elements $g_{R,L}$ are now $x$-dependent.

The field strengths corresponding to the external potentials are given as
usual by
\be
F_{R,L}^{\mn}=\dmu(v^\nu\pm a^\nu)-\dnu(v^\mu\pm a^\mu)-
i[v^\mu\pm a^\mu,v^\nu\pm a^\nu].
\ee
The covariant derivatives of $U$ and $R_\mu$ are given by
\be
D_\mu U=\dmd U-i(\vpamu)U+iU(\vmamu),
\ee
and
\be
\nabla_\mu R_\nu=\dmd R_\nu+[\Gamma_\mu,R_\nu]
\ee
with the connection 
\be
\Gamma_\mu=\frac{1}{2}\left(u^\dag[\dmd-i(\vpamu)]u+u[\dmd-i(\vmamu)]u^\dag
\frac{}{}\right)
\ee
Thus the building elements of the effective Lagrangian are  
$U$, $D_\mu U$, $F_{R,L}^{\mn}$, $R_\mu$ and $\nabla_\mu R_\nu$.

Note that while some of the above variables ($U$, $D_\mu U$, $F_{R,L}^{\mn}$)
transform under the full group $G$, others ($R_\mu$ and $\nabla_\mu R_\nu$)
transform under the unbroken subgroup $H$. One may take advantage of the 
mixed transformation rule of $u$ to redefine the former variables so 
as to transform under $H$ only. Thus one introduces the field variables
\cite{Ecker89}, 
\be
u_\mu=iu^\dag D_\mu U u^\dag =u_\mu^\dag
\ee
and
\be
f_\pm^{\mn}=\pm u^\dag F_R^{\mn} u + u F_L^{\mn} u^\dag
\ee

We now use these variables to write down the leading terms of the 
different pieces of the Lagrangian density, which are invariant under $H$ 
and hence also under $G$. To calculate the correlation functions of
isotriplets of vector and axial-vector currents, we need the Lagrangian
for the interacting system of both the isotriplets and isosinglets of vector
and axial-vector mesons and pions in presence of the external fields. (We
shall see later that other particles cannot contribute to the pole terms to
order $T^2$.) For the pions we have the familiar term,
\be
\cl(\pi)=\frac{\F^2}{4}\la u_\mu u^\mu \ra .
\ee
Here and below the symbol $\la A \ra$ stands for the trace of the 2$\times$2
matrix $A$. For the spin-1 meson isotriplet and isosinglet fields, let
\be
R_{\mn}=\nabla_\mu R_\nu-\nabla_\nu
R_\mu~~,~~~~S_{\mu\nu}=\dmd S_\nu -\dnd S_\mu .
\ee
Then the kinetic terms take the form
\be
\cl_{kin}(R,S)=\sum\left( -\frac{1}{4}\la R_{\mn} R^{\mn}\ra+
\frac{1}{2}m_R^2\la R_\mu R^\mu \ra \right)
+\sum\left( -\frac{1}{4}S_{\mn} S^{\mn} +\frac{1}{2} m_S^2 S_\mu
S^\mu\right),
\ee
where the two sums run over the isotriplets and the isosinglets respectively. 

The leading coupling terms linear in  the octets of the vector and the 
axial-vector fields have been obtained in Ref.\cite{Ecker89,Ecker89a} for 
the symmetry group $SU(3)\times SU(3)$. As mentioned already,
the symmetry reduces to  $SU(2)\times SU(2)$ at finite temperature. In
effecting this reduction, we keep the masses of the physical particles in each  
of the octets to be the same but consider the interaction separately for the 
isotriplets $ [\rho (770), a_1 (1230)]$ and isosinglets 
$[\om (782), f_1 (1282)]$. Restricting to terms 
relevant to one loop calculations, we get for the vector meson couplings,
\be
\cl_{int}(V)=\frac{1}{2\rt m_V}( \Fr\la\rho_{\mn}\fmn_+ 
\ra+iG_\rho\la\rho_{\mn}[u^\mu,u^\nu]\ra  
+ iH_\rho\la \rho_\mu[u_\nu,\fmn_- ]\ra +H_\om \emn \om^\mu \la u^\nu
f_+^{\lm\sg}\ra ),
\ee
while for the axial vector meson such couplings are
\be
\cl_{int}(A)=\frac{1}{2\rt m_A}( F_{a_1}\la a_{1\,\mn}\fmn_- 
\ra+iH_{a_1}\la a_{1\,\mu}[u_\nu,\fmn_+ ]\ra 
+H_{f_1}\emn f_1^{\mu}\la u^\nu f_-^{\lm\sg}\ra ).
\ee
Finally we write down the quadratic couplings of the triplets with the
singlets and between themselves,
\be
\cl_{int}(V,A)= \frac{1}{\rt} \emn (g_1 \om^\mu\la \rho^{\nu\lm}u^\sg
\ra + g_2 f_1^\mu\la a_1^{\nu\lm}u^\sg\ra ) 
+ \frac{i}{2}g_3\la \rho^{\mn}[a_{1\,\mu},u_\nu]\ra .
\ee
(The term $\la a_1^{\mn}[\rho_\mu,u_\nu]\ra $ is equivalent to the third term
above to leading order in pion momentum.) Although we have here a number 
of different coupling constants, only some of them will actually enter our 
results for the shifts in particle parameters to $O(T^2)$.

\section{Perturbative vertices}
\setcounter{equation}{0}
\renewcommand{\theequation}{3.\arabic{equation}}
  
The vertices needed in evaluating the one loop Feynman diagrams for the
correlation functions can be recognised best from the diagrams themselves. 
Let us consider the correlation function of two vector currents at non-zero
temperature $T=1/\bet$,
\be
i\int d^4x\, e^{iq\cdot x}\,Tr\,\varrho\,T\,V_\mu^a(x)\,V_\nu^b(0)
~~,
\ee
where $\varrho = e^{-\bet H}/Tr\,e^{\bet H} $ is the thermal density matrix
of QCD. The diagrams contributing to it are of three types, 
namely those of self-energy, vertex modification and intermediate states
shown respectively in Figs.~1, 2 and 3. We see that we only need vertices up to 
four fields, counting both quantum (particle) and classical (external) fields. 
At each vertex the pion field can appear at most quadratically. When it does
appear so, the two pion fields must be contracted at the same vertex. The
resulting pion loop is of $O(T^2)$, if no derivative is present on the pion
fields; but the loop is of higher order if it does, allowing us to ignore such
vertices. Also at each vertex the external fields can occur at most linearly 
(with the exception of the vertex in Fig.~3(b)). At such a vertex with an 
external field, the vector or the axial vector meson field must also occur
linearly. Keeping these requirements in mind, we now obtain the 
polynomial version of the chiral Lagrangian written in the previous section.

\bef[t]
\centerline{\psfig{figure=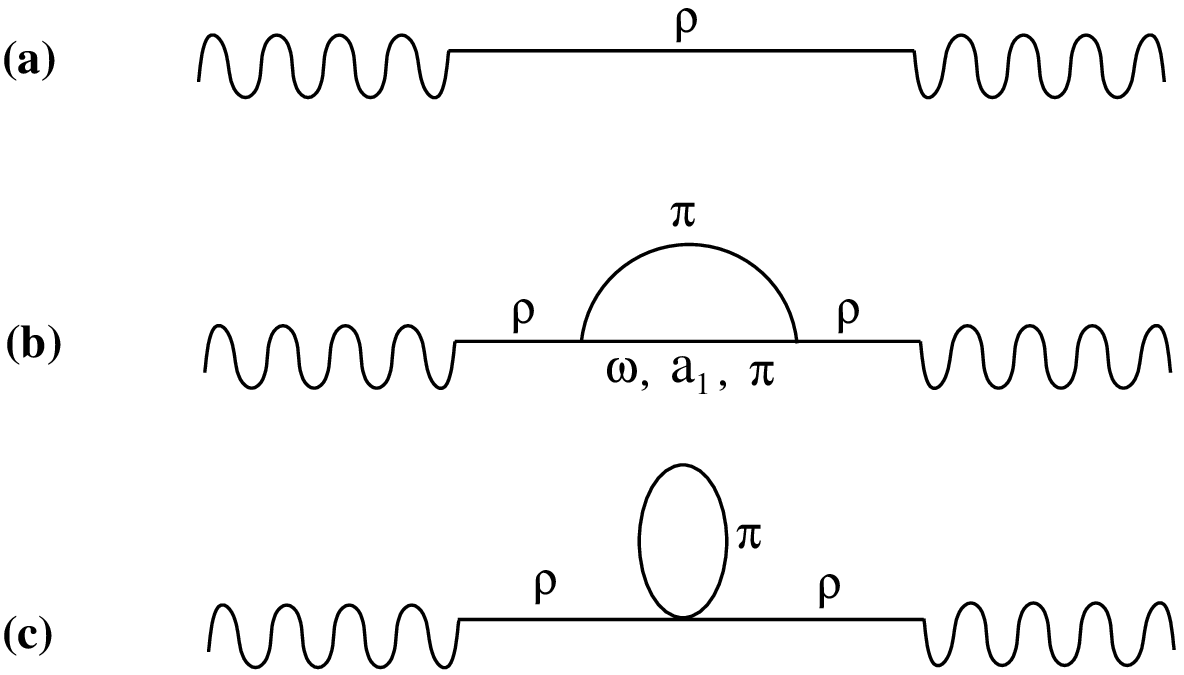,height=5.5cm,width=9cm}}
\caption{ $\rho$ meson pole and self-energy diagrams}
\eef

\bef[b]
\centerline{\psfig{figure=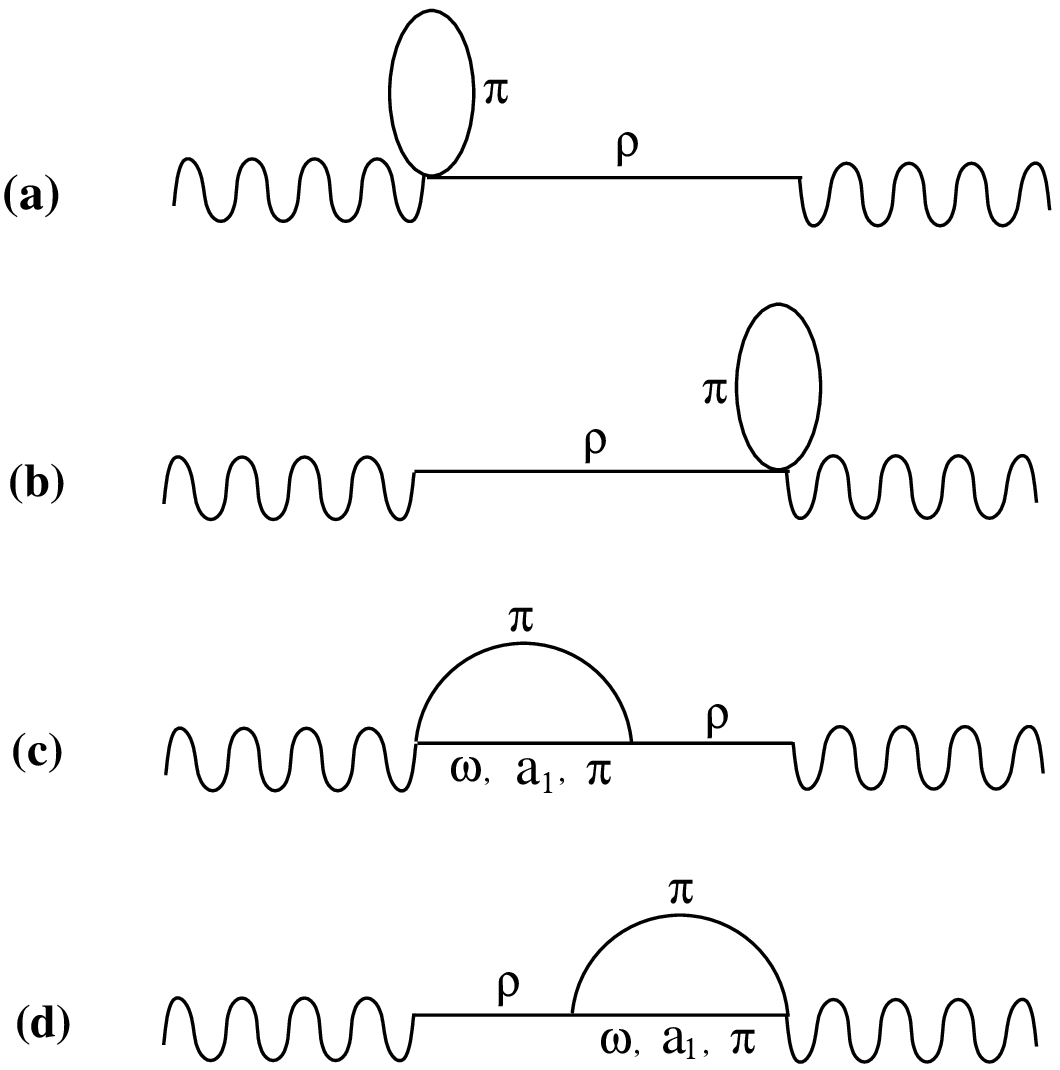,height=8cm,width=9cm}}
\caption{Vertex correction diagrams}
\eef

\bef
\centerline{\psfig{figure=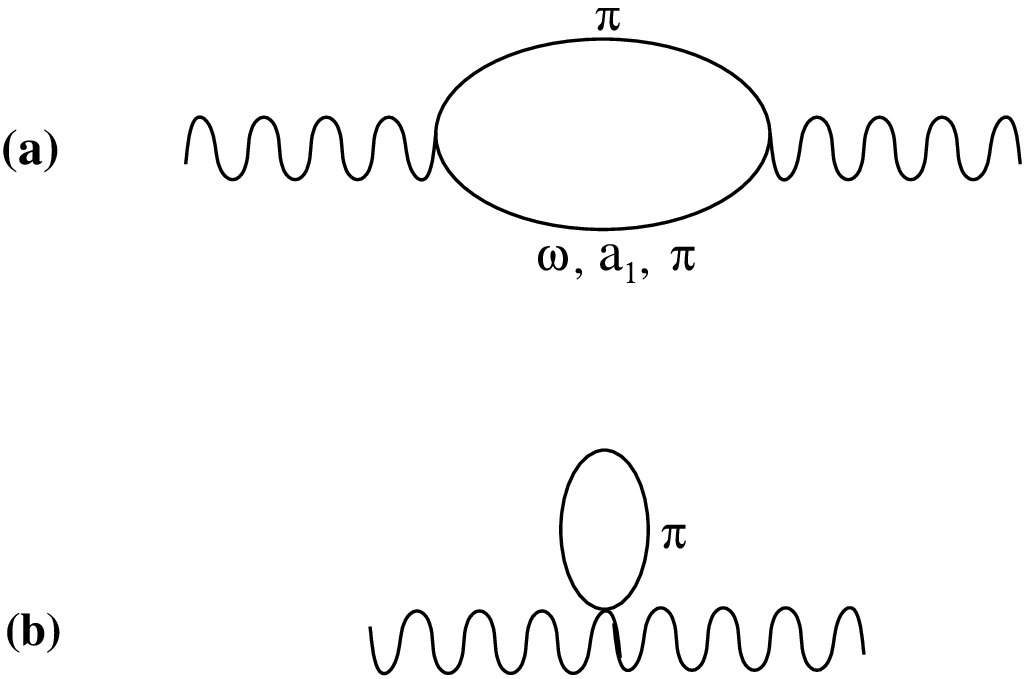,height=5.5cm,width=9cm}}
\caption{Intermediate state diagrams}
\eef

The pion Lagrangian (2.17) can be expanded as 
\bea
\cl(\pi)&=& \frac{1}{2}(\dmd\bp\cdot\dmu\bp-m_\pi^2\bp\cdot\bp)
-\F\ba_\mu\cdot\dmu\bp + \bv_\mu\cdot\bp\times\dmu\bp
+\F\bp\cdot(\bv_\mu\times\ba^\mu) \nonumber \\
&+& \frac{1}{2}\F^2\ba_\mu\cdot\ba^\mu
+\frac{1}{2}\left\{\bp\cdot\bp (\bv_\mu\cdot\bv^\mu-\ba_\mu\cdot\ba^\mu )
-\bv_\mu\cdot\bp \bv^\mu\cdot\bp + \ba_\mu\cdot\bp\ba^\mu\cdot\bp \right\}.
\eea
Here and below we denote the isovector triplet of fields by bold face 
letters. Note that we have added the pion mass term, even though we shall 
evaluate the final results in the chiral limit. The first term gives the 
free, thermal pion propagator matrix. As discussed in the Appendix, all
we need here is its 11-component, 
\bea
\int &&d^4x\, e^{ik\cdot x}\,Tr\,\varrho\,T\pi^a(x)\pi^b(0) |_{11}\nonumber\\
&&=\delta^{ab}\left( \frac{i}{k^2-m_\pi^2+i\ep}+2\pi n(k_0)
\delta(k^2-m_\pi^2)\right),
\eea
where  $n$ is the bosonic distribution function
$n(k_0)=(e^{\beta|k_0|}-1)^{-1}$.

The kinetic part of the Lagrangian for the spin one mesons reduces to
\be
\cl_{kin} (R,S)=-\frac{1}{2}\sum_{\br,\ba_1} \{ \dmd\R_\nu\cdot
(\dmu\R^\nu - \dnu\R^\mu )-m_R^2\R_\mu\cdot\R^\mu\} -
\frac{1}{2}\sum_{\om,f_1}\{\dmd S_\nu(\dmu S^\nu -\dnu
S^\mu) -m_S^2 S_\mu S^\mu\}.
\ee
The density of spin one mesons are suppressed exponentially in the heat 
bath. So we need only the vacuum part of the respective 11-components 
of the propagator matrices, which for $R_\mu^a$, for example, is 
\be
\int d^4x\, e^{ip\cdot x}\,\la 0|T R_\mu^a(x)R_\nu^b(0)|0\ra
=\de^{ab}\frac{i(-g_{\mu\nu}+p_\mu p_\nu /m_R^2)}{p^2-m_R^2+i\ep}~.
\ee

As an illustration we now work out the vertices present in the term 
$\la\rho_{\mn}\fmn_+ \ra$. Since
$\fmn_+$ is proportional to the external fields, we need only the terms in 
$\rho_{\mn}$ without them. Thus upto four field vertices, we have
\bea
\la\rho_{\mn}\fmn_+ \ra&=&4\la\dmu v^\nu(\dmd \rho_\nu-\dnd \rho_\mu)\ra
+\frac{2i}{\F}\la\dmu a^\nu[\pi,\dmd \rho_\nu-\dnd \rho_\mu]\ra\nonumber\\
&&-\frac{1}{2\F^2}\la\dmu v^\nu[\pi,[\pi,\dmd \rho_\nu-\dnd \rho_\mu]]\ra
\eea
As already explained, we have to contract the two pion fields in the four 
field vertex above. Its thermal piece is
\be
Tr\,\varrho\,T\pi^a(x)\pi^b(x) |_{11} \rw 
\de^{ab}\int\,\frac{d^4k}{(2\pi)^3}n(k_0)\delta(k^2-m_\pi^
2)=\de^{ab}\frac{T^2}{12}, 
\ee
in the chiral limit. (We see here that any derivative on the pion fields
would result in contributions of higher order in $T$). Thus the third term
in Eq.(3.6) reduces to the first with a $T$-dependent coefficient.

In this way we derive the pieces of the polynomial Lagrangian from those of the
chiral Lagrangian of Eqs. (2.20-22) as
\bea
\cl (V)&=&\frac{F_{\rho}}{m_V}\left\{\left(1-\frac{T^2}{12\F^2}\right)\dmu\bv^\nu\cdot
(\dmd\br_\nu-\dnd\br_\mu) 
+ \frac{1}{\F}\dmu\ba^\nu\cdot(\dmd\br_\nu-\dnd\br_\mu)\times{\bp}
\right\}\nonumber \\
& &-\frac{2G_{\rho}}{m_V\F^2}\dmd\br_\nu\cdot\dmu\bp\times\dnu\bp 
-\frac{\rt H_\om}{m_V\F}\emn \om^{\mu}\dnu\bp\cdot
\partial^\lm\bv^\sg ,
\eea
\bea
\cl (A)&=&-\frac{F_{a_1}}{m_A}\left\{\left(1-\frac{T^2}{12\F^2}\right)
\dmu\ba^\nu\cdot(\dmd\ba_{1\nu}-\dnd\ba_{1\mu}) 
+ \frac{1}{\F}\dmu\bv^\nu\cdot(\dmd\ba_{1\nu}-\dnd\ba_{1\mu})\times{\bp}
\right\}\nonumber \\
& & +\frac{\rt H_{f_1}}{ m_A\F}\emn f_1^{\mu}\dnu\bp\cdot
\partial^\lm\ba^\sg,
\eea
and $[u\overleftrightarrow\dmd v\equiv (\dmd u) v-u\dmd v]$
\be
\cl (V,A)=\emn\left( \frac{g_1}{\F}\om^\mu\overleftrightarrow\dnu 
\br^\lm\cdot\partial^\sg\bp +\frac{g_2}{\F} 
 f_1^\mu\overleftrightarrow\dnu  \ba_1^\lm\cdot\partial^\sg\bp\right) +
\frac{g_3}{\F}
\dmu\br^\nu\cdot(\ba_{1\,\mu}\times\dnd\bp-\ba_{1\,\nu}\times\dmd\bp).
\ee
In writing the above expressions, when two or more vertices present
themselves with the same field content, differing only in the number of
derivatives on the pion field, we retain only the one with fewer or no
derivative on it.  

The coupling constants can generally be determined from the observed decay
rates of the particles. Thus the decay rate $\Gamma (\rho^0
\rw e^+e^-)=6.9 $ keV gives $F_{\rho} =154$ MeV. Similarly the decay
rate $\Gamma (a_1\rw \pi\gamma)= 640$ keV gives $F_{a_1} =135 $ MeV. 
The latter constant can also be determined from one of the Weinberg sum
rules \cite{Weinberg}, agreeing closely with this value.
The decay width $\Gamma (\rho\rw 2\pi) =153$ MeV gives the coupling
$G_{\rho} = 69$ MeV. Here one expects large chiral corrections as the pions
are far from being soft. Thus the chiral loops reduce this value to
$G_{\rho} = 53$ MeV \cite{Ecker89}. Using this value and the decay rate 
$\Gamma (\om\rw 3\pi)=7.6$ MeV, we get the dimensionless coupling
constant $g_1 =0.87$ \cite{Gell-Mann}. There does not appear any
data relating to the decay of $f_1$ to determine $g_2$. The remaining
coupling constants in Eqs. (3.8-10) will not appear in our results to
$O(T^2)$.

\section{Mass and coupling shifts} 
\setcounter{equation}{0}
\renewcommand{\theequation}{4.\arabic{equation}}

Given the vertices, it is simple to calculate the diagrams of Figs.~1-3. The
Lorentz tensor and the thermal matrix structures of the two point functions 
are discussed in the Appendix, according to which the free pole term of 
Fig.~1(a) is given in the variable $E\equiv q_0$ for $\vq =0 $ essentially by
\[-E^4\cdot\frac{(F_\rho/m_V)^2}{E^2-m_V^2+i\ep}~.\]

We now determine how this pole position and the residue
are modified by interactions at finite temperature. The self-energy diagrams 
of Fig.~1 modify it to
\[
-E^4\cdot\frac{(F_\rho/m_V)^2}{E^2-m_V^2-\Pi_t}
\]
where we use Eqs. (A.12-13) to construct $\Pi_t$ from the finite temperature 
part of the polarisation tensor $\Pi_{\al\bet}$,
\be
\Pi_{\al\bet}(q)=\sum_{c=\om,a_1,\pi}\Pi_{\al\bet}^{(c)}(q)~~,
\ee
the sum running over the diagrams of Fig.~1(b). (We have left out Fig. 1(c),
which is clearly of order $T^4$ .)  To write out these
contributions we define the gauge invariant tensors
\bea
A_{\al\bet}(q)&=&-g_{\al\bet}+{q_\al q_\bet}/{q^2} ,\nonumber\\
B_{\al\bet}(q,k)&=&q^2 k_\al k_\bet-q\cdot k(q_\al k_\bet+k_\al q_\bet)
+(q\cdot k)^2g_{\al\bet} ,\nonumber\\
C_{\al\bet}(q,k)&=&q^4 k_\al k_\bet-q^2(q\cdot k)(q_\al k_\bet+k_\al q_\bet)
+(q\cdot k)^2q_\al q_\bet .
\eea
The three contributions are now given by
\be
\Pi_{\al\bet}^{(\om)}= - \left(\frac{2g_1}{\F}\right)^2
\int\frac{d^4k}{(2\pi)^3}\,\frac{n(k_0)\,\delta(k^2-m_\pi^2)}
{(q-k)^2-m_V^2}\,(q^2k^2A_{\al\bet}+B_{\al\bet})\, ,
\ee
\be
\Pi_{\al\bet}^{(a_1)}=-2\left(\frac{g_3}{\F}\right)^2 
\int\frac{d^4k}{(2\pi)^3}\,\frac{n(k_0)\,\delta(k^2-m_\pi^2)}
{(q-k)^2-m_A^2}\,(B_{\al\bet}-C_{\al\bet}/m_A^2)\, ,
\ee
and
\be
\Pi_{\al\bet}^{(\pi)}=2\left(\frac{2G_\rho}{m_V\F^2}\right)^2
\int\frac{d^4k}{(2\pi)^3}\,\frac{n(k_0)\,\delta(k^2-m_\pi^2)}
{(q-k)^2-m_\pi^2}\,C_{\al\bet}\, .
\ee
The respective contributions to $\Pi_t$ are ($\omk=\sqrt{\vk^2+m_\pi^2}$),
\be
\Pi_t^{(\om)}(E)=\frac{16g_1^2}{3\F^2}\,E^2(E^2-m_V^2+m_\pi^2)
\int\frac{d^3k\, n(\omk)}{(2\pi)^3\,2\omk}\cdot\frac{\vec k^2}
{(E^2-m_V^2+m_\pi^2)^2-4E^2\omk^2}\, ,
\ee
\be
\Pi_t^{(a_1)}(E)=\frac{4}{3}\left(\frac{g_3}{\F}\right)^2
E^2(E^2-m_A^2+m_\pi^2)\int\frac{d^3k\, n(\omk)}{(2\pi)^3\,2\omk}\cdot
\frac{\vec k^2 (2+E^2/m_A^2)+3m_\pi^2 }
{(E^2-m_A^2+m_\pi^2)^2-4E^2\omk^2}\, ,
\ee
and
\be
\Pi_t^{(\pi)}(E)=\frac{4}{3}\left(\frac{2G_\rho}{m_V\F^2}\right)^2 
E^6\int\frac{d^3k\, n(\omk)}{(2\pi)^3\,2\omk}\cdot\frac{\vec k^2}
{E^4-4E^2\omk^2}\, .
\ee

To evaluate the integrals for $E^2$ near $m_V^2$ in the chiral limit, we
note the difference in the small $\vec{k}$ behaviour of the denominators of 
the integrands: while it behaves like $\sim \vec{k}^2$ in Eq.~(4.6), 
they are constants in Eq.~(4.7-8). Accordingly we get \cite{Comment},
\be
\Pi_t^{(\om)}(E)=-\frac{g_1^2T^2}{18\F^2}(E^2-m_V^2),\;
~~~\Pi_t^{(a_1)}\sim\Pi_t^{(\pi)}\sim O(T^4)~.
\ee
Including also the constant vertex corrections from Figs. 2(a,b), we get
for the complete pole term
\be
-E^4\frac{(F_\rho^T/m_V)^2}{E^2-m_V^2}
\ee
with
\be
F_\rho^T=F_\rho\left\{1-\left( 1+ \frac{g_1^2}{3} \right)\frac{T^2}{12\F^2}
\right\}.
\ee

It is simple to see that the remaining diagrams cannot alter this pole
term to $O(T^2)$. Each of the remaining vertex corrections of Fig. 2(c,d) is
essentially of the form of the corresponding self-energy integral $(\Pi_t)$ 
multiplied by the $\rho$-meson pole. Then the behaviour (4.9) of the 
self energies
excludes any pole contribution to $O(T^2)$ from these corrections. Of the
contributions from the diagrams of Fig. 3 with intermediate states, the one
with $\pi\om$ is like $\Pi_t^{(\om)}$, while those with $\pi a_1$ and $\pi\pi$
have two powers of $\vk$ fewer in the integrands compared to the self energies
$\Pi_t^{(a_1)}$ and $\Pi_t^{(\pi)}$ respectively. Clearly they cannot contribute
to the $\rho$ meson pole. 

It is now also simple to establish that any particle with a mass other than
$m_V$ and not included in our calculation, cannot
alter our results. The argument rests simply on the structure of the interaction
vertices without external fields. Such a vertex must have a pion field with
a derivative. Thus the loop integrals involving such a particle must be of
order higher than $T^2$.

It is interesting to work out the amplitude of Fig. 3(a) corresponding to the
intermediate state $\pi a_1$.
Its finite temperature part is given by 
\be
-2\left(\frac{F_{a_1}}{m_A\F}\right)^2\int\frac{d^4k}{(2\pi)^3}
\frac{n(k_0)\,\delta (k^2-m_\pi^2)}{(q-k)^2 -m_A^2} \{q^2(q^2-2q\cdot
k)A_{\mn} -B_{\mn}\}.
\ee
It may be readily evaluated in the chiral limit to give essentially 
\cite{Comment}
\be
\frac{T^2}{6\F^2}\left\{
-E^4\frac{(F_{a_1}/m_A)^2}{E^2-m_A^2}\right\} ,
\ee
where the expression in second bracket is the contribution of the 
axial vector meson $a_1$ to the vacuum
correlation function of axial vector currents.
Observe the different origin of the $O(T^2)$ correction to the $\rho$ and
$a_1$ poles. While for $\rho$, it is given by the self-energy due to 
$\pi\omega$ and constant vertex diagram, it is given by the intermediate 
state $\pi a_1$ for $a_1$. This presence of {\it axial} vector meson 
pole in the {\it vector} current correlation function at finite temperature 
was detected earlier using PCAC and current algebra \cite{Dey}. 

One can calculate the axial-vector meson pole term in the axial-vector
current correlation function in an entirely analogous manner. Again one
finds that the pole position does not shift from $E^2=m_A^2$, and the
residue $F_{a_1}^T$ is given by Eq. (4.11) with $g_1$ replaced by $g_2$.

\section{Discussion}

Here we have determined the temperature dependence of the parameters of
$\rho$ and $a_1$ meson poles appearing in the vector and the axial-vector
correlation functions. It is based on the flavour symmetry of QCD and its
spontaneous breakdown, as embodied in chiral perturbation theory. It allows
a low temperature expansion of these quantities, of which we just
calculate the leading term to $O(T^2)$. To this order the mass shift is
zero, due to the presence of derivatives on pion fields in interaction 
vertices as required by the chiral symmetry. 
The residues depend on the coupling constants at the vertices $\rho\om\pi$ 
and $a_1f_1\pi$ and decrease with temperature. These results parallel those for
the nucleon calculated earlier, the above vertices playing roles analogous 
to that of the vertex $\pi N\bar{N}$ \cite{Leutwyler90}.

We may try to estimate the range of validity of our results by comparing
these with the temperature expansions of other quantities. For the quark
condensate the expansion has been worked out to order $T^6$~\cite{Gasser87}. 
It shows that the $T^2$ term continues to dominate the series up to $T=150$ 
MeV. It may not be true in all cases, however. Take the example of the nucleon
pole residue in the nucleon correlation function. In the pion-nucleon system
there is the low energy resonance $\Delta (1237)$, which limits considerably
the range of validity of the first few terms in the chiral perturbation
series for $\pi N$ scattering amplitudes. Already at $T=100$ MeV, the average 
pion energies in the heat bath are outside this range. This situation
limits the validity of the $T^2$ term in the nucleon pole residue to
temperatures below $100$ MeV. 

As the $\pi\rho$ and $\pi a_1$ systems do not have any such low energy
excitations, we may expect our calculation to order $T^2$ to represent 
the expansion well up to temperatures of about $150 $ MeV, as in the case of the
quark condensate. There is, however, one source of enhanced chiral corrections
to our results, which may limit this range of validity. As mentioned earlier, 
at the vertices where the $\rho$ or the external fields couple to two pions 
(see Figs. 1(b), 2(c,d) and 3(a)), the pion momenta are not at all soft. Thus,
although these diagrams do not contribute to order $T^2$, the coefficients
of higher order terms may be larger.

We finally comment briefly on the different approaches in the literature to 
determine the temperature dependence of these parameters. First, there are
other Lagrangian approaches. One has the massive Yang-Mills \cite{Meissner}
and the hidden gauge \cite{Bando} formulations. Since these Lagrangians are
built on the $SU(2)\times SU(2)$ flavor symmetry, they also produce no shift
in the meson masses to $O(T^2)$ \cite{Harada,Song}. However, the 
coefficients of $T^4$ quoted in these works must be incomplete, as they 
do not include two loop diagrams. One has also the more phenomenological
Lagrangian of Quantum Hadrodynamics \cite{Serot}, which produces negligible 
shift in the meson masses to $O(T^2)$~\cite{Alam}.

The second approach tries to calculate the correlation functions using soft
pion techniques, but fails to reproduce the terms proportional to $g_1^2$ and 
$g_2^2$ in $F_\rho$ and $F_{a_1}$ respectively \cite{Dey}. Since chiral 
perturbation theory actually incorporates such techniques in a systematic 
and refined way, it is interesting to trace this disagreement. The Tr(ace) 
in Eq. (3.1) can be restricted to the vacuum and the one pion states, 
giving the full contribution to $O(T^2)$ in the chiral limit. In the pion 
matrix element one may use the hypotheses of partially conserved axial-vector 
current and the algebra of currents to arrive at
\[
Tr \,\varrho T\,V_\mu^a (x)V_\nu^b (0) 
=\left( 1-\frac{T^2}{6\F^2} \right) \la 0|T\,V_\mu^a(x) V_\nu^b
(0)|0\ra_\beta
+\frac{T^2}{6\F^2} \la 0 |T\,A_\mu^a (x) A_\nu^b (0)|0\ra ,
\]
and a similar one for the axial vector correlation function. The subscript
$\beta$ in the first matrix element on the right is a reminder that 
it is really not a vacuum matrix element, since any internal (pion) line
which would occur in its perturbative evaluation must be taken as thermal.
Being multiplied with $T^2$ , the second matrix element can be taken as a
true vacuum expectation value for results to $O(T^2)$ and we have already 
verified the correctness of this term in our calculations above.
 
The authors of Ref.\cite{Dey} assume the first matrix element also to be a
true vacuum matrix element  having no temperature dependence.
In terms of Feynman diagrams it amounts to ignoring the self
energy diagrams of Fig. 1, which can in general contribute to both the pole
position and the residue. In fact, the missing $g_1^2$ term in 
$F_\rho $ arises exactly from the self energy diagram of Fig. 1(b).
A similar situation arises also in the case of the correlation function of
nucleon currents. Applying the same soft pion techniques and assuming the
resulting matrix elements to be true vacuum expectation values, one misses
the term proportional to the axial coupling constant of the nucleon in the
residue of the nucleon pole \cite{Koike}. We conclude that such soft pion
techniques as applied to the correlation functions cannot reproduce the full
results of chiral perturbation theory. 

The third approach is based on thermal QCD sum rules \cite{Hatsuda,Mallik98}. 
It is convenient to discuss these sum rules after subtracting out the 
corresponding vacuum sum rules, leaving only $T$ dependent contributions. 
The spectral side of such a sum rule is given by three types of one loop 
diagrams, like those of Fig.1, 2 and 3. In the literature, however, it is 
assumed to be saturated by a pole term with $T$ dependent parameters and 
the two particle intermediate states containing (at least) one pion. The 
inadequacy of this saturation scheme is revealed by
the fact that the self-energy  and vertex correction diagrams are generally
not exhausted by contributions which make the pole term $T$ dependent. It is
the neglect of the remaining contributions which make the existing results
unreliable. (For incorporation of such remaining contributions in the
nucleon sum rule, see Refs. \cite{Koike,Mallik01a}.)

The fourth and the last one is a phenomenological approach based on the 
(first order) virial expansion of the self energy of a particle
\cite{Leutwyler90,Mallik01}. Quite generally it expresses the
complex shift in the pole position of a particle as an integral over 
the product of the forward scattering amplitude of a pion on the particle 
and the pion distribution function. Since the pion gas is rather dilute even
at a temperature $\sim 100$ MeV, this formula should hold upto higher
temperatures than the results of chiral perturbation expansion to first
order only. But the problem here is to construct the scattering amplitude, 
which is reliable \cite{Eletsky01}.

\section*{Appendix}
\setcounter{equation}{0}
\renewcommand{\theequation}{A.\arabic{equation}}

Here we discuss the kinematic structure of the vector and axial-vector 
two point functions. In the
real time formulation of thermal field theory we are using here, one
has with each physical vertex an associated (ghost) vertex to take into
account \cite{Niemi}. This causes all two point functions to take the form of 
2$\times$2 matrices. Thus even if we begin with the 11-component, the 
perturbative expansion will mix it up with the other components. But as we 
show below, this complication does not arise for the problem at hand and we
actually need work with the 11-component only.

Consider first the complete vector (or axial-vector) meson propagator, which
appears in the self-energy diagrams of Fig.1. Denoting the thermal 
$2\times2$ matrices 
in this Appendix by bold face letters, it satisfies the Dyson equation,
\be
\D_{\mn}=\D_{\mn}^{(0)}+\D_{\mu\lm}^{(0)}
(-i{\bP}^{\lm\sigma}) \D_{\sigma\nu}
\ee
The free thermal propagator has the factorised form
\begin{equation}
\D_{\mn}^{(0)}(q_0,\vec q)
= \U(q_0)~\left( 
\begin{array}{cc} 
D_{\mn}^{(0)} & 0 \\
0 & D_{\mn}^{(0)\,\ast}
\end{array} \right)\;
\U(q_0)
\ee
where
\be
\U(q_0)  = \left( \begin{array}{cc} \sqrt{1+n} & \sqrt{n} \\
\sqrt{n} & \sqrt{1+n}
\end{array} \right)~~~,~~~n=(e^{\beta|q_0|}-1)^{-1}
\ee
and ${D}_{\mn}^{(0)}(q)$ is the vacuum propagator
\be
D^{(0)}_{\mn}(q)=\frac{i(-g_{\mn}+q_\mu q_\nu/m_R^2)}{q^2-m_R^2+i\ep}
\ee
From its spectral representation, the complete propagator $i\D_{\mn}(q)$
can also be shown to admit a similar factorisation,
\be
\D_{\mn}(q_0,\vec q)
= \U(q_0)~\left( \begin{array}{cc} 
D_{\mn} & 0 \\
0 & D_{\mn}^{^\ast}
\end{array} \right) \U(q_0)
\ee
It then follows from Eq.~(A1) that $\U(-i\bP_{\mn})\U$ must have the
diagonal structure,
\be
\U(-i\bP_{\mn})\U=\left( \begin{array}{cc}
-i\Pi_{\mn} & 0 \\
0 & i\Pi_{\mn}^* \\
\end{array}\right)
\ee
The matrix equation (A.1) now collapses to the ordinary equation
\be
{D}_{\mn}={D}_{\mn}^{(0)}+{D}_{\mu\lm}^{(0)}
(-i{\Pi}^{\lm\sigma}) {D}_{\sigma\nu}
\ee
and its complex conjugate. The only remnants of the matrix structure  
are the relations between $\Pi_{\mn}$ and the components of $\bP_{\mn}$. 
With the 11-component of the latter, these are
\be
Re(\bP_{11})_{\mn}=Re\Pi_{\mn}~~~,~~~Im(\bP_{11})_{\mn}=(1+2n)Im\Pi_{\mn}.
\ee

In calculations at finite temperature, one usually prefers the matter rest 
frame, thereby losing explicit Lorentz covariance. The latter may be restored 
by bringing in the four velocity $u_\mu$ in the medium \footnote{No
confusion should arise from the earlier use of $u_\mu$ as a field variable
in sec. II}. Then the time and space
components of a four-vector $q_\mu$ are converted to Lorentz scalars,
$\om=u\cdot q$ and $\bq=\sqrt{\om^2-q^2}$. In this framework a gauge covariant
decomposition of the polarisation tensor is
\be
\Pi_{\mn}(q)=\Pi_t\,P_{\mn}+\Pi_l\,Q_{\mn},
\ee
where we choose the kinematic covariants as
\be
P_{\mn}=-g_{\mn}+\frac{q_\mu q_\nu}{q^2}-\frac{q^2}{\bq^2}\tilde u_\mu\tilde
u_\nu~~,~~~~~~Q_{\mn}=\frac{q^4}{\bq^2}\tilde u_\mu\tilde u_\nu~~,
\ee
with $\tilde u_\mu=u_\mu-\om q_\mu/q^2$.
The invariant amplitudes are functions of two variables, say $q^2$and $\om$.
With the decomposition (A.9) the Dyson equation (A.7) can be solved to get
\be
D_{\mn}=\frac{i}{q^2-m_R^2-\Pi_t+i\ep}P_{\mn}
+\frac{i}{q^2-m_R^2-q^2\Pi_l +i\ep}\frac{Q_{\mn}}{q^2}
+\frac{i}{m_R^2}\,
\frac{q_\mu q_\nu}{q^2}~.
\ee

To find the invariant amplitudes $\Pi_{t,l}$ we form the scalars
\be
\Pi_1=\Pi^\mu_\mu~~,~~~~\Pi_2=u^\mu u^\nu \Pi_{\mn}~~,
\ee
when Eq. (A.9) gives
\be
\Pi_l=\frac{1}{\bq^2}\Pi_2~~~,~~~~~\Pi_t=-\frac{1}{2}(\Pi_1+\frac{q^2}{\bq^2}
\Pi_2)~.
\ee
In the limit $\vec q\rightarrow 0$,the kinematic structures (A.10 ) 
relate the two invariant amplitudes,
\be
\Pi_t(q_0,\vec q=0)=q_0^2\Pi_l(q_0,\vec q=0)~.
\ee
In this limit the second equation of (A.13) simplifies to
\be
\Pi_t=-\frac{1}{3}\Pi_1 .
\ee

Having discussed the 2$\times$2 matrix structure of the complete
propagator, we turn to the same for the two point functions, 
which for definiteness we take the one for vector currents. From the 
self-energy diagrams of Fig.~1 it gets the contribution
\bea
&&\de^{ab}\left(\frac{F_V}{m_V}\right)^2\,(q^2g_\mu^\lm-q_\mu q^\lm)\,\D_{\lm
\sg}\,(q^2g_\nu^\sg-q^\sg q_\nu)\nonumber\\
&&=\de^{ab}\U(q_0)\,
\left(
\begin{array}{cc}
G_{\mn} & 0 \\
0 & -G_{\mn}^{\ast}
\end{array} \right)\;
\U(q_0)
\eea
where
\be
G_{\mn}=- q^4\left(\frac{F_\rho}{m_V}\right)^2 \left(\frac{1}
{q^2-m_V^2-\Pi_t+i\ep}P_{\mn}
+\frac{1}{q^2-m_V^2-q^2\Pi_l+i\ep}\frac{Q_{\mn}}{q^2}\right)
\ee
on using (A.11). Note that at $\vq =0$ the two pole positions above coincide
due to the kinematic constraint (A.14).
The vertex correction diagrams of Figs.~2(c) and (d) are of the form
\be
\bP^{\mu\lm}\D_{\lm\nu}^{(0)}=\U^{-1}(q_0)~\left(
\begin{array}{cc}
\Pi^{\mu\lm}D_{\lm\nu}^{(0)} & 0 \\
0 & \Pi^{\mu\lm\,\ast}D_{\lm\nu}^{(0)\,\ast}
\end{array} \right)\;
\U(q_0)
\ee
while the diagrams Fig.~3(a) with intermediate states are of the form given by
(A.5).

We see that the results of evaluation of different diagrams are typically
of the form of a diagonal matrix, sandwiched between $\U$ or $\U^{-1}$.
Near the meson pole the  $\U$ matrix reduces to the unit
matrix with exponential correction $\sim e^{-m_\rho/T}$.
Such corrections are too small to be retained  when we are calculating 
corrections of $O(T^2)$ only. The remaining diagonal matrix may be 
represented by its 11-component, which again admits a decomposition 
into invariant amplitudes like that of $\Pi_{\mn}$ in Eq.~(A.9) with the
longitudinal and the transverse ones being again related for $\vq =0$ as 
in Eq. (A.14) for $\Pi_{l,t}$.

\section*{Acknowledgement}

One of us (S.M.) acknowledges support of CSIR, Government of India.

\end{document}